\documentclass[a4paper]{jpconf}
\usepackage{graphicx}
\usepackage{bm}

\begin{document}

\title{Recent progresses on the pseudospin symmetry in single particle resonant states}

\author{Bing-Nan Lu$^{1}$, En-Guang Zhao$^{1,2}$ and Shan-Gui Zhou$^{1,2}$}
 \address{$^1$ State Key Laboratory of Theoretical Physics,
               Institute of Theoretical Physics, Chinese Academy of Sciences, 
               Beijing 100190, China}
 \address{$^2$ Center of Theoretical Nuclear Physics, National Laboratory
               of Heavy Ion Accelerator, Lanzhou 730000, China}
 \ead{sgzhou@itp.ac.cn (Shan-Gui Zhou)}

\begin{abstract}
The pseudospin symmetry (PSS) is a relativistic dynamical symmetry 
directly connected with the small component of the nucleon Dirac wave function. 
Much effort has been made to study this symmetry in bound states.
Recently, a rigorous justification of the PSS in single particle 
resonant states was achieved by examining the asymptotic behaviors
of the radial Dirac wave functions:
The PSS in single particle resonant states 
in nuclei is conserved exactly when the attractive scalar and repulsive vector 
potentials have the same magnitude but opposite sign.
Several issues related to the exact conservation 
and breaking mechanism of the PSS in single particle resonances
were investigated by employing spherical square well potentials 
in which the PSS breaking part can be well isolated in the Jost function.
A threshold effect in the energy splitting and 
an anomaly in the width splitting of pseudospin partners
were found 
when the depth of the square well potential varies from zero to a finite value.
\end{abstract}

\section{Introduction}

In 1969, the pseudospin symmetry (PSS) was found in atomic nuclei:
Doublets of single particle levels with quantum numbers
$(n_{r},l,j=l+1/2)$ and $(n_{r}-1,l+2,j=l+3/2)$ in the same major shell 
are nearly degenerate~\cite{Arima1969_PLB30-517,Hecht1969_NPA137-129}.
Since the PSS was observed, 
much effort has been devoted to investigate its origin~\cite{Bohr1982_PS26-267, 
Dudek1987_PRL59-1405,
Castanos1992_PLB277-238, 
Bahri1992_PRL68-2133, 
Blokhin1995_PRL74-4149}.
Finally, 
it was revealed that the PSS in nuclei is a relativistic symmetry
which is exactly conserved when the scalar potential $S(r)$
and the vector potential $V(r)$ have the same size but opposite sign, 
i.e., $\Sigma(r)\equiv S(r)+V(r)=0$~\cite{Ginocchio1997_PRL78-436},
or, more generally, when $d\Sigma(r)/dr=0$~\cite{Meng1998_PRC58-R628, Meng1999_PRC59-154}.
The PSS is usually viewed as a dynamical symmetry~\cite{Arima1999_RIKEN-AF-NP-276,
Alberto2001_PRL86-5015} because,
in either limit, $\Sigma(r)=0$ or $d\Sigma(r)/dr=0$, there are no bound states 
in realistic nuclei, thus in realistic nuclei the PSS is always broken.

The SS and PSS have been investigated extensively.
The readers are referred to Ref.~\cite{Ginocchio2005_PR414-165} 
for a review and to Refs.~\cite{Liang2013_PRC87-014334, Lu2013_PRC88-024323} 
for overviews of recent progresses.
Here we simply mention that 
the study of the PSS has been extended along several lines:
\begin{enumerate}
\item
From single particle states in the Fermi sea to those in the
Dirac sea, i.e., the spin symmetry in anti-nucleon spectra~\cite{Ginocchio1999_PR315-231, 
Zhou2003_PRL91-262501, He2006_EPJA28-265, Liang2010_EPJA44-119}.
\item
From normal nuclei to hypernuclei~\cite{Song2009_CPL26-122102, 
Song2011_CPL28-092101, Song2010_ChinPhysC34-1425}.
\item
From spherical systems to deformed ones~\cite{Lalazissis1998_PRC58-R45, 
Sugawara-Tanabe1998_PRC58-R3065, 
Ginocchio2004_PRC69-034303,
Guo2013_arXiv1309.1523}.
\item
From bound states to resonant states; next we will focus on this extension.
\end{enumerate}

There has been a great interest in the exploration 
of continuum and resonant statesin atomic nuclei~\cite{Yang2001_CPL18-196,
Zhang2003_SciChinaG46-632,Meng2006_PPNP57-470}.
The study of symmetries, e.g., the PSS, in resonant states is very interesting.
There have been some numerical investigations of the PSS in single particle 
resonances~\cite{Guo2005_PRC72-054319, Guo2006_PRC74-024320, Liu2013_PRA87-052122, 
Zhang2006_HEPNP30S2-97, Zhang2007_CPL24-1199} and
the SS in single particle resonant states~\cite{Xu2012_IJMPEE21-1250096}.
Recently, we gave a rigorous justification of the PSS in single particle 
resonant states and investigated several open problems 
related to the exact conservation and breaking mechanism of 
the PSS in single particle resonances~\cite{Lu2012_PRL109-072501, Lu2013_AIPCP1533-63, 
Lu2013_PRC88-024323}.
In this contribution, we will present some results concerning the PSS
in single particle resonant states.

\section{Justification of the PSS in resonant states}

In atomic nuclei, the equation of motion for nucleons, the Dirac equation, reads,
\begin{equation}
 \left[ \bm{\alpha}\cdot\bm{p} + \beta(M+S(\bm{r}))+V(\bm{r}) \right]\psi(\bm{r})
 = \epsilon \psi(\bm{r}),
~\label{Eq:Diraceq}
\end{equation}
where $\bm{\alpha}$ and $\beta$ are Dirac matrices, $M$
is the nucleon mass, and $S(\bm{r})$ and $V(\bm{r})$ the scalar and vector potentials. 

For a spherical nucleus, the upper and lower components of the Dirac spinor are
$i {F_{n\kappa}(r)}/{r} \cdot Y_{jm}^{l}(\theta,\phi)$ and 
$- {G_{\tilde{n}\kappa}(r)}/{r} \cdot Y_{jm}^{\tilde{l}}(\theta,\phi)$, respectively.
One can derive a Schr\"odinger-like equation for the lower component, 
\begin{eqnarray}
 \left[
   \frac{d^{2}}{dr^{2}}
  -\frac{1}{M_{-}(r)}\displaystyle\frac{d\Sigma(r)}{dr}\frac{d}{dr}
  -\displaystyle\frac{\tilde{l}(\tilde{l}+1)}{r^{2}}
  + \displaystyle\frac{1}{M_{-}(r)}
    \frac{\kappa}{r}\displaystyle\frac{d\Sigma(r)}{dr}
  -  M_+(r) M_-(r)
 \right] G(r)
 & = & 
 0 , 
 \label{eq:G}
\end{eqnarray}
where $M_+(r)=M+\epsilon-\Delta(r)$ and $M_-(r)=M-\epsilon+\Sigma(r)$.

For the continuum in the Fermi sea, $\epsilon\geq M$; 
the regular solution for Eq.~(\ref{eq:G}) vanishes at the origin
%
and at large $r$,
\begin{equation}
 G(r) = \frac{i}{2}\left[\mathcal{J}^G_{\kappa}(p)     h_{\tilde{l}}^{-}(pr) - 
                         \mathcal{J}^G_{\kappa}(p)^{*} h_{\tilde{l}}^{+}(pr) \right],
 \ \ r\rightarrow\infty,
\label{Eq:Jost_Gdef}
\end{equation}
where $h_{\tilde{l}}^{\pm}(pr)$ are the Ricatti-Hankel functions.
$\mathcal{J}_{\kappa}(p)$ is the Jost function 
which is an analytic function of $p$.
The zeros of $\mathcal{J}_{\kappa}^{G}(p)$ 
on the lower $p$ plane and 
near the real axis correspond to resonant states. 
The resonance energy $E$ and width $\Gamma$ are determined by the relation 
$E-i{\Gamma}/{2}=\epsilon - M$. 

In the PSS limit, Eq.~(\ref{eq:G}) is reduced as
\begin{equation}
 \left[
          \frac{d^{2}}{dr^{2}}
        - \frac{\tilde{l}(\tilde{l}+1)}{r^{2}}
        + \left(
                \epsilon-M
          \right)
          M_+(r)
 \right]  G(r)
 =  0
 .
~\label{eq:PSLeqG}
\end{equation}
For pseudospin doublets with $\kappa$ and 
$\kappa^{\prime}=-\kappa+1$, the small components satisfy the same equation because 
they have the same pseudo orbital angular momentum $\tilde{l}$.
For continuum states, $G_{\kappa}(\epsilon,r)=
G_{\kappa^{\prime}}(\epsilon,r)$ is true for any energy $\epsilon$ thus 
we have $\mathcal{J}_{\kappa^{\prime}}^{G}(p)=\mathcal{J}_{\kappa}^{G}(p)$. 
Thus the zeros are the same for $\mathcal{J}_{\kappa^{\prime}}^{G}(p)$ and 
$\mathcal{J}_{\kappa}^{G}(p)$ and the PSS in single particle resonant states 
in nuclei is exactly conserved~\cite{Lu2012_PRL109-072501}.

\section{Numerical Results and Discussions}

Spherical square well potentials for $\Sigma(r)$ and $\Delta(r)$ read,
\begin{equation}
 \Sigma(r) = \left\{ \begin{array}{c}
                         C,\qquad r<R,\\
                         0,\qquad r\geq R,
                         \end{array}\right.
 \qquad
 \Delta(r) = \left\{ \begin{array}{c}
                         D,\qquad r<R,\\
                         0,\qquad r\geq R,
                         \end{array}\right.
 \label{eq:S.W.}
\end{equation}
where $C$ and $D$ are depths and $R$ is the width. 
The Jost function $\mathcal{J}_{\kappa}^{G}(p)$ is derived as~\cite{Lu2012_PRL109-072501},
\begin{equation}
 \mathcal{J}_{\kappa}^{G}(p) =  
 - \frac{p^{\tilde{l}}}{2ik^{\tilde{l}+1}}
   \left\{ j_{\tilde{l}}(kR) p h_{\tilde{l}}^{+\prime}(pR)
        -kj_{\tilde{l}}^{\prime}(kR) h_{\tilde{l}}^{+}(pR)
    - b
     \left[ kj_{\tilde{l}}^{\prime}(kR) - \frac{\kappa}{R} j_{\tilde{l}}(kR) \right] 
     h_{\tilde{l}}^{+}(pR)
   \right\}, 
   \nonumber \\ 
 \label{eq:Jost_function}
\end{equation}
with $b=     {C}/{(\epsilon-M-C)}$ and 
$k = \sqrt{\left( \epsilon - C - M \right) \left( \epsilon - D + M \right)}$.
By examining the zeros of the Jost function (\ref{eq:Jost_function}), one gets 
single particle bound states and resonant states.

\begin{figure}
\begin{centering}
\includegraphics[width=0.6\textwidth]{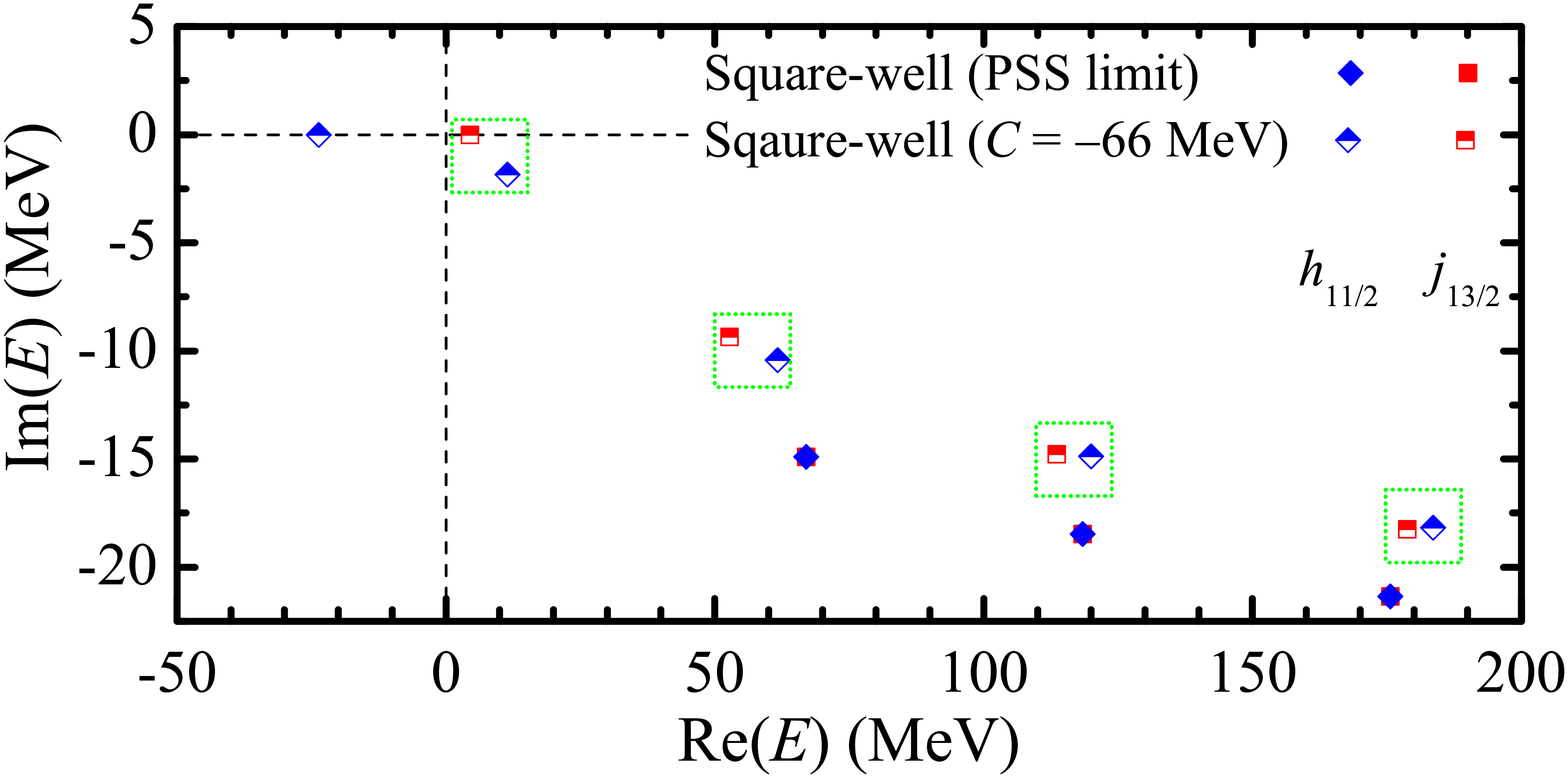}
\par\end{centering}
\caption{(Color online)
The zeros of the Jost function $\mathcal{J}_{\kappa}^{G}$ on the complex energy plane
in square-well potentials (\ref{eq:S.W.}) with 
$C=0$ (solid symbols) and $C=-66$ MeV (half-filled symbols) 
for PS partners $h_{11/2}$ (diamond) and $j_{13/2}$ (square). 
}
~\label{fig:zeros}
\end{figure}

Figure~\ref{fig:zeros} shows solutions for PSS doublets with $\tilde{l}=6$, 
i.e., $h_{11/2}$ with $\kappa = -6$ and $j_{13/2}$ with $\kappa' = 7$ 
for square-well potentials with $D=650$ MeV and $R=$ 7 fm.
The conservation of the PSS for single particle resonant states is seen:
When $C=0$, the roots for $h_{11/2}$ and $j_{13/2}$ are the same. 
When $C=-66$ MeV, the PSS is broken.

\begin{figure}
\begin{centering}
\includegraphics[width=0.40\columnwidth]{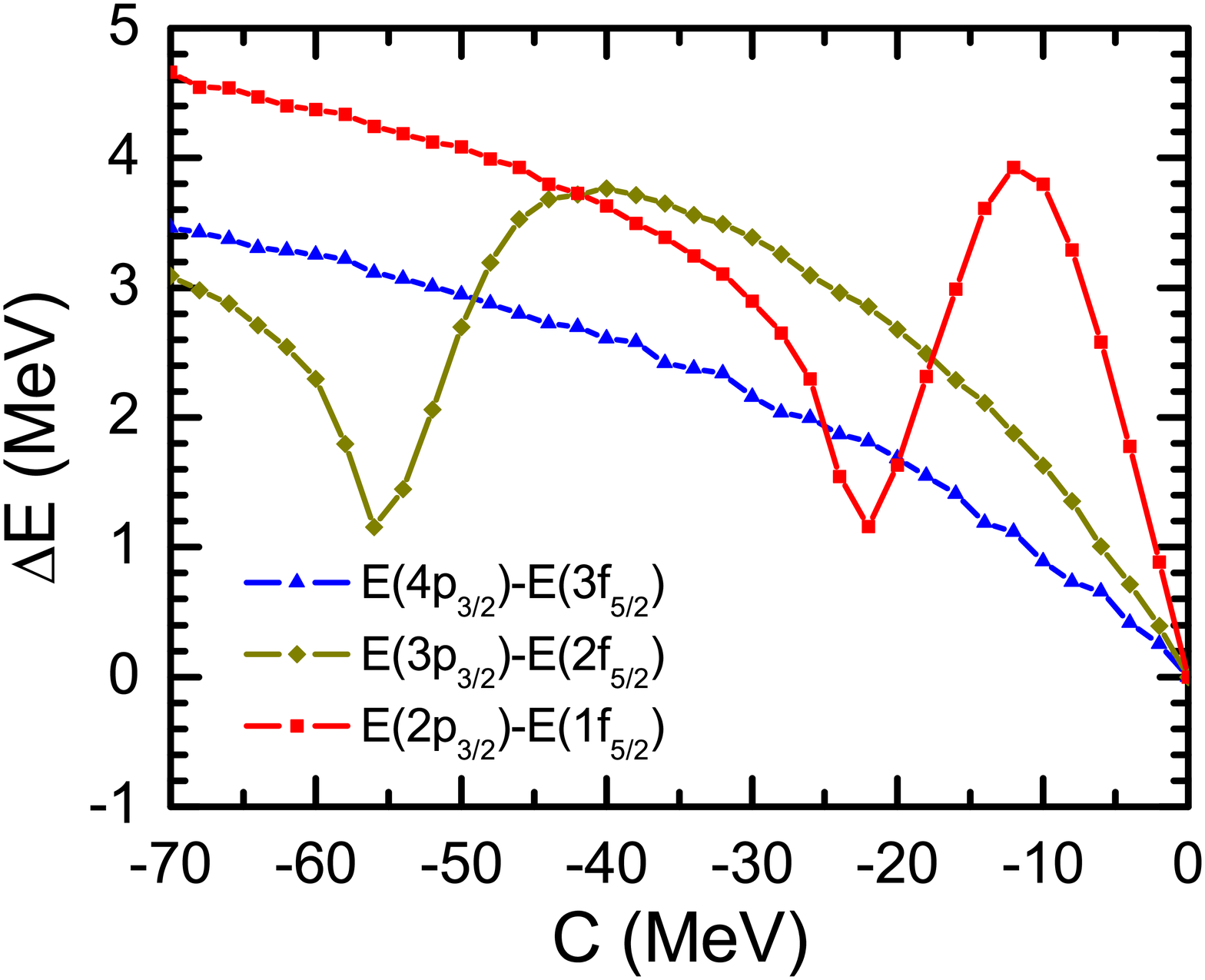}
\includegraphics[width=0.40\columnwidth]{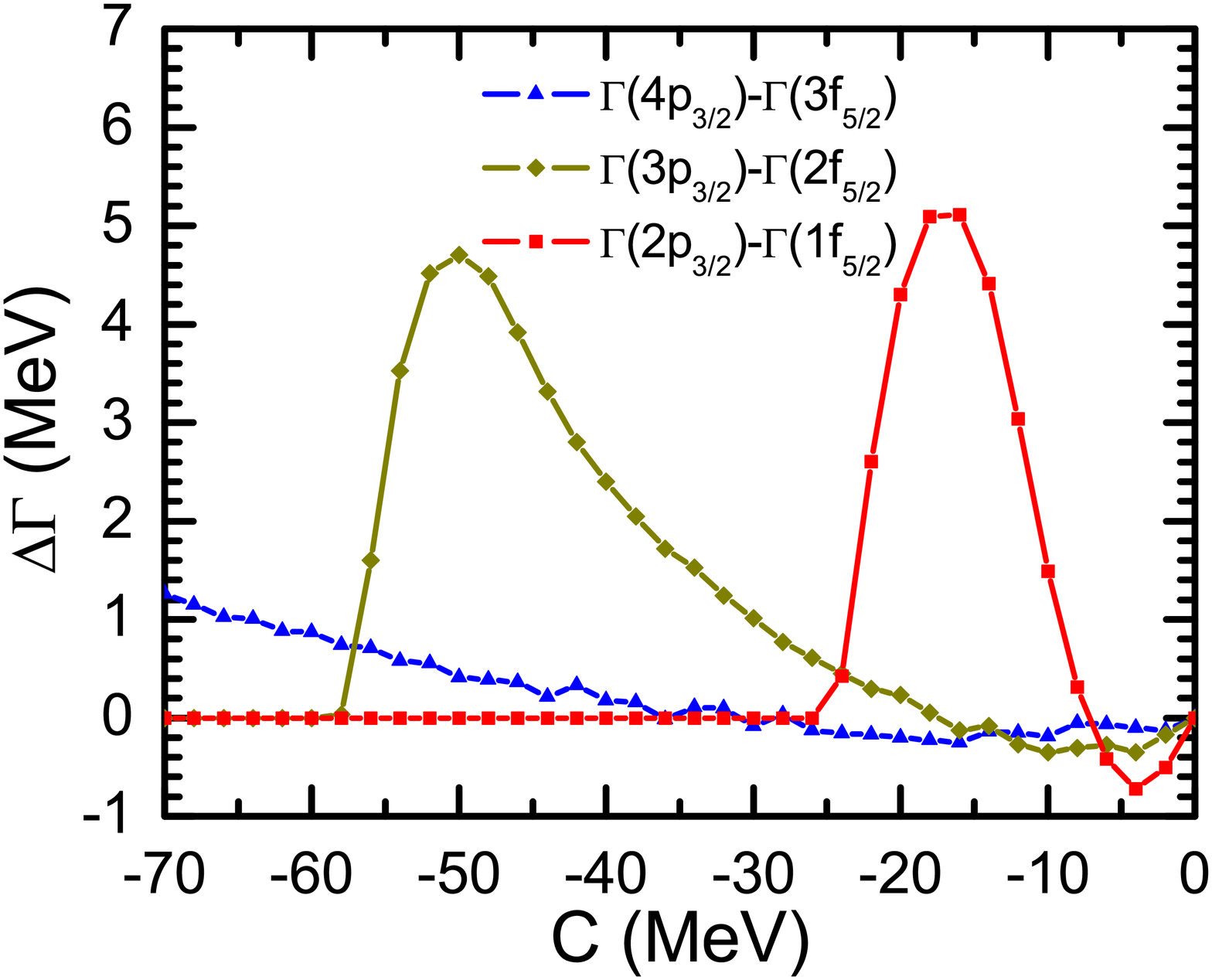}
\par\end{centering}
\caption{\label{fig:Delta_EG}(Color online).
The energy (the left panel) and width (the right panel) 
splitting between PSS partners with the pseudo orbital 
angular momentum $\tilde{l}=2$, $p_{3/2}$ and $f_{5/2}$,  
as a function of the potential depth $C$.
Taken from Ref.~\cite{Lu2013_PRC88-024323}.
}
\end{figure}

By examining the Jost function, one can trace continuously the PSS partners 
from the PSS limit to the case with a finite potential depth. 
This has been done in Ref.~\cite{Lu2013_PRC88-024323} 
in which 
we found a threshold effect in the energy splitting and 
an anomaly in the width splitting of pseudospin partners 
when $C$ varies from zero to a finite value.
As the depth of the single particle potential becomes deeper, 
the PSS is broken more and  
a threshold effect in the energy splitting appears which
can be seen in Fig.~\ref{fig:Delta_EG}:
The energy splitting first increases then decreases until the pseudospin doublets encounter 
the threshold where one of the levels becomes a bound state
and the splitting takes a minimum value; 
When the potential becomes even deeper, the splitting increases again.
When the depth of the single particle potential increases from zero,
there also appears an anomaly in 
the width splitting of PS partners which is shown in
Fig.~\ref{fig:Delta_EG}: It first decreases from zero to a maximum value with
negative sign, then increases and becomes zero again; 
after the inversion of the width splitting, the splitting increases
and reaches a maximum and positive value, then it becomes smaller and finally reaches zero.

Finally we mention that there are still many open problems 
concerning the exact and breaking of the PSS in single particle resonant states; 
the readers are referred to Refs.~\cite{Lu2013_PRC88-024323, Lu2013_AIPCP1533-63}
for further discussions.

\ack 
This work has been supported by Major State Basic Research Development 
Program of China (Grant No. 2013CB834400), 
National Natural Science Foundation of China (Grant Nos. 11121403, 11175252, 
11120101005, 11211120152, and 11275248), 
the Knowledge Innovation Project of Chinese Academy of Sciences (Grant No. KJCX2-EW-N01). 

\section*{References}


\providecommand{\newblock}{}

\end{document}